\documentclass[useAMS,usenatbib]{mn2e}
\usepackage{url,times,graphicx,amsmath,amsfonts,amssymb,aas_macros,color,epsfig,varioref,subfigure,comment,natbib}
\usepackage{color}
\newcommand{\scb}{\textcolor{black}}

\def\be{\begin{equation}}
\def\ee{\end{equation}}
\def\bi{\begin{itemize}}
\def\ei{\end{itemize}}
\def\tt{\texttt}

\def\aap{A\&A}

\def\apjl{APJL}

\newcommand{\Tab}[1]{Table~\ref{#1}}
\newcommand{\Sec}[1]{Section~\ref{#1}}
\newcommand{\App}[1]{Appendix~\ref{#1}}
\newcommand{\Eq}[1]{Eq.~(\ref{#1})}
\newcommand{\Fig}[1]{Fig.~\ref{#1}}
\newcommand{\hMpc}{{\ifmmode{h^{-1}{\rm Mpc}}\else{$h^{-1}$Mpc}\fi}}
\newcommand{\hGpc}{{\ifmmode{h^{-1}{\rm Mpc}}\else{$h^{-1}$Gpc}\fi}}
\newcommand{\hkpc}{{\ifmmode{h^{-1}{\rm kpc}}\else{$h^{-1}$kpc}\fi}}

\newcommand{\hMsun}{{\ifmmode{h^{-1}{\rm {M_{\odot}}}}\else{$h^{-1}{\rm{M_{\odot}}}$}\fi}}
\newcommand{\Msun}{{\ifmmode{{\rm {M_{\odot}}}}\else{${\rm{M_{\odot}}}$}\fi}}
\def\hMpc{$h^{-1}\,{\rm Mpc}$}
\def\hkpc{$h^{-1}\,{\rm kpc}$}

\def\kms{{ \rm km} $s^{-1}$}
\def\LCDM{\ensuremath{\Lambda}CDM}
\def\vtan{$v_{\rm tan}$}
\def\vtanI{$v_{\rm tan}^{(I)}$}
\def\vtanII{$v_{\rm tan}^{(II)}$}
\def\vrad{$v_{\rm rad}$}
\def\mlg{$M_{\rm LG}$}
\def\cde{cDE}
\def\frfou{$\rm FofR04$}
\def\frfiv{$\rm FofR05$}
\def\frsix{$\rm FofR06$}
\def\dgpot{DGP12}
\def\dgpfs{DGP56}
\def\symma{SymmA}
\def\symmb{SymmB}

\def\modz{Mod0}
\def\modo{Mod1}
\def\modtw{Mod2}
\def\modth{Mod3}

\def\mvir{$M_{\rm vir}$}

\def\ered{$e_{\rm red}$}
\def\lorb{$l_{\rm orb}$}
\def\mmw{$M_{\rm MW}$}
\def\mmto{$M_{\rm M31}$}
\def\mlg{$M_{\rm LG}$}

\title
[The dynamics of the local group as a probe of Dark Energy and Modified Gravity]
{The dynamics of the local group as a probe of Dark Energy and Modified Gravity}
\author[Carlesi, Mota \& Winther]
{
Edoardo Carlesi $^{1}$,
David F. Mota $^2$,
Hans A. Winther $^{3,4}$
\\
\thanks{E-mail: carlesi@phys.huji.ac.il}
$^{1}$Racah Institute of Physics, Givat Ram, 91040 Jerusalem, Israel\\
$^2$ Institute for Theoretical Astrophysics, University of Oslo, Norway\\
$^3$ Astrophysics,  University of  Oxford, DWB, Keble  Road, Oxford,
  OX1 3RH, United Kingdom\\
$^4$ Institute of Cosmology and Gravitation, University of Portsmouth,Burnaby
Road, Portsmouth PO1 3FX, United Kingdom
}
\begin{document}

\date{Submitted XXXX XXX XXXX}

\pagerange{\pageref{firstpage}--\pageref{lastpage}} \pubyear{2016}

\maketitle

\label{firstpage}


\begin{abstract}
In this work we study the dynamics of the Local Group (LG) within the context of cosmological models beyond General Relativity (GR).
Using observable kinematic quantities to identify candidate pairs we build up samples of simulated LG-like
  objects drawing from $f(R)$, symmetron, DGP and quintessence {\it N}-body simulations together with their \LCDM\ counterparts
featuring the same initial random phase realisations.
The variables and intervals used to define LG-like objects are referred to as Local Group model;
different models are used throughout this work and adapted to study their dynamical and kinematic properties.
The aim is to determine how well the observed LG-dynamics can be reproduced within cosmological theories beyond GR, 
We compute kinematic properties of samples drawn from alternative theories and \LCDM\ and
compare them to actual observations of the LG mass, velocity and position.
As a consequence of the additional pull, pairwise tangential and radial velocities are enhanced in modified gravity and coupled dark energy with respect to \LCDM\, 
inducing significant changes to the total angular momentum and energy of the LG.
For example, in models such as $f(R)$ and the symmetron 
this increase can be as large as $60\%$, peaking well outside of the $95\%$ confidence region allowed by the data.
This shows how simple considerations about the LG dynamics can lead to clear small-scale observational signatures 
for alternative scenarios, without the need of expensive high-resolution simulations.
\end{abstract}

\begin{keywords}
Cosmology, Numerical simulations, Dark matter, Local Group
\end{keywords}

\section{Introduction}\label{sec:intro}

The accelerated expansion of the Universe still remains largely unexplained
since its discovery at the turn of the century \citep{Riess:1998, Perlmutter:1999}.
One way to account for it is adding an extra component with negative equation of state to the
General Relativity (GR) Lagrangian, the so called dark energy (DE). 
In its simplest form, DE takes the form of a cosmological constant ($\Lambda$) which
together with the Cold Dark Matter (CDM) paradigm defines the current standard concordance model \LCDM.
Despite its simplicity and its successes, a number of theoretical problems \citep[see][for a comprehensive review]{Bull:2016}
have led the theorists to devise models of dynamical dark energy, 
with a time-dependent equation of state, such as quintessence \citep{Wetterich:1995, Caldwell:1998,
Copeland:1998, Zlatev:1999},
vector dark energy \citep{BeltranMaroto:2008, Carlesi:2012}, $\kappa -$essence \citep{Picon:2000} and Chaplygin gas \citep{Kamenshchik:2001} to mention some
among the great number of models elaborated during the last years.

However, DE (in both its static and dynamic form) is not the only possible explanation for the late time acceleration of the Universe:
an alternative mechanism for this involves
large scales modifications of GR \citep[see ][for a comprehensive review]{Clifton:2012}.
This latter class of theories usually introduces an additional scalar degree of freedom alongside the usual massless spin-2
graviton of GR, which effectively acts as a fifth force.
In general, in order to keep this class of theories consistent with local GR tests \citep{Will:2014},
it is necessary to introduce some kind of interaction-screening within high-density regions \citep{2010arXiv1011.5909K}.\\

Cosmological simulations of dark energy models \citep{Baldi:2012de}
and modified gravity \citep{Winther:2015} have emerged over the last decade as the main tool to study several aspects of the
non-linear regime of theories beyond \LCDM.
For instance, they have been used to look for signatures in the
matter power spectrum and mass functions \citep{Maccio:2004, Baldi:2010a, Cui:2012, Viel:2012, 2012JCAP...01..051L, 2013MNRAS.436..348P, mota, He:2015, mota2,VargasDosSantos:2016},
in voids, environment and the cosmic web \citep{Li:2011, Winther:2012, Shim:2014, Falck:2014, Carlesi:2014a, Elyiv:2015, Sutter:2015, Pollina:2016},
and in the properties of galaxy clusters \citep{Lee:2012, Llinares:2013, Carlesi:2014b, Hammami:2015, Gronke:2015, 2014MNRAS.440..833A}.\\

Model selection among the great number of alternatives to the standard paradigm,
devising new tests that might help constrain their free parameters and their consistency, is a fundamental task of theoretical cosmology.
In this work we introduce a new way of estimating the viability of several types of modified gravity and dark energy models,
which employs {\it N}-body simulations and uses the observed dynamics of the Local Group (LG) of galaxies as a cosmological probe.
We focus on the properties of its two most prominent members, the Milky Way (MW) and Andromeda (M31) spirals,
to determine whether alternative theories can account for their observed kinematics. 
This is a computationally cheap and conceptually simple way of using astrophysical scales, cosmology-independent data as a test for theories beyond GR.
In fact, since in such an approach halos are treated as point-like
particles, disregarding the details of inner structure, 
one does not have to rely on expensive high-resolution simulations to deal with
the sub-megaparsec regime of alternative cosmologies.\\

We focus here on several models where a fifth-force kind of interaction
is introduced: coupled quintessence \citep{Amendola:2000}, $f(R)$ \citep{hu}, the symmetron \citep{hinter} and
DGP \citep{2000PhLB..485..208D}.
In these cases, deviations from standard Newtonian gravity and
the enhancement of particles' accelerations 
\citep[a common feature of most fifth-force models, see e.g.][]{Baldi:2010a,Gronke:2015,Shi:2015}
is expected to affect the two body dynamics of the LG in a systematic way, providing clear observational
signatures \citep{Hellwing:2014}.
The aim of the present work is to quantify these signatures and compare them to the observations, 
in order to determine which models (or parameters) turn out to be at odds (or in agreement) with the known LG dynamics.
This is achieved by computing posterior distribution functions (PDF) for a series of dynamic (energy and momentum) and kinematic
(radial and tangential velocity components) variables, using samples of LG-like objects found in the simulations.
To reduce the arbitrariness in the definition of a \emph{simulated} Local Group, we employ several 
definitions using different observationally-motivated parameters and intervals, which we refer to as \emph{models} of the LG.
The PDFs are then computed for each LG model and each simulation.
Comparing these functions to the actual values, we can establish to which extent a cosmology is expected to account
for the observed LG dynamics. \\

Of course, one may ask about the statistical relevance of such a procedure given that we have data related to only \emph{one} LG.
Doing cosmology with a single object however is not a new challenge, and despite its limitations, it has been successfully employed e.g.  
in the case of massive high-$z$ galaxy clusters \citep{Baldi:2010, Carlesi:2011,Harrison:2011, Wazimann:2011} and
the bullet cluster \citep{Lee:2012}, where single, peculiar objects have been used to evaluate whether their
existence could be considered a normal outcome of the model or else was indicating some inconsistency.
Moreover, the idea is in line with the concept of \emph{near-field cosmology} \citep{Bland-Hawthorn:2006}, that aims to
extract cosmologically relevant informations from the study of nearby objects, under the assumption that
our patch of the Universe is not an extraordinary environment but rather a common kind of place. 
This assertion could be of course contradicted invoking some kind of anthropic principle, arguing that
our galaxy is an outlier and a one-of-a-kind object inside our Universe.
Rejecting the latter hypothesis, however, it becomes meaningful to analyse the properties of MW size objects  
for cosmological purposes, as it has been done in recent years to argue against \LCDM\ \citep{Kroupa:2012, Pawlowski:2015}, 
in favour of it \citep{Libeskind:2015a}, or to compare it to alternative models
\citep{Penzo:2014, Elahi:2015, Maccio:2015, Penzo:2016}.\\

This paper is structured as follows. \Sec{sec:models} briefly introduces the main motivations, mathematical and
physical features that characterize the different class of models (quintessence, symmetron and $f(R)$) that will
be discussed later.
\Sec{sec:simulations} contains a description of the simulations, the parameters used
and the basic properties of the non-standard {\it N}-body codes used.
In \Sec{sec:methods} we discuss the main ideas behind our formalism and the properties of the observational data it relies upon.
The results of our analysis are then presented in \Sec{sec:results}, where we discuss the implications for the
viability and non viability of some of the models presented here.
A summary of the techniques and the results discussed throughout the \emph{Paper} is presented in \Sec{sec:conclusions}.


\section{Models}\label{sec:models}
This section provides a short introduction to the main physical and mathematical properties
of the models discussed. References to more accurate descriptions are 
provided in each subsection for the interested reader.

\subsection{DGP}

DGP  \citep{2000PhLB..485..208D} is  a so-called braneworld  model where  matter
lives on a 4D brane which is embedded in a 5D spacetime. The action is
given by
\begin{align}
S    =    \int\sqrt{-g_{(4)}}{\rm d}^4x\frac{M_{\rm    Pl~(4)}^2}{2}R_{(4)}    +
\int\sqrt{-g_{(5)}}{\rm d}^5x\frac{M_{\rm Pl~(5)}^2}{2}R_{(5)}
\end{align}

where $g_{(4)}$ denotes the induced  4D metric on the brane, $R_{(4)}$
the induced  Ricci scalar on  the brane,  $g_{(5)}$ the metric  in the
bulk and $R_{(5)}$ the Ricci scalar in the bulk.

The     ratio    of     the    two     Planck    masses,     $r_c    =
\frac{1}{2}\left(\frac{M_{\rm  Pl~(4)}}{M_{\rm Pl~(5)}}\right)^2$,  is
the only free parameter of the model known as the crossover scale. For
scales $r \ll r_c$  gravity behaves as being four dimensional while for
$r\gtrsim r_c$  the five dimensional aspects become important.

The modifications  to gravitational  force is  determined by  a scalar
field $\phi$ called the brane-bending mode. The brane-bending mode influences the dynamics of particles through a gravitational potential
\begin{align}
\Phi = \Phi_N + \frac{\phi}{2}
\end{align}
where $\Phi_N$ is the standard Newtonian potential, i.e. $\nabla^2\Phi_N = 4\pi Ga^2\delta\rho_m$. The dynamics of
$\phi$ in the quasi-static approximation \citep{2015PhRvD..92f4005W} is given by
\begin{align}
\nabla^2\phi   +   \frac{r_c^2}{3\beta  a^2}\left[(\nabla^2\phi)^2   -
  (\nabla_i\nabla_j\phi)^2\right] =  \frac{a^2\delta\rho}{\beta M_{\rm
    Pl}^2}
\end{align}
where
\begin{align}
\beta(a) = 1 + 2H(a)r_c\left(1 + \frac{\dot{H}(a)}{3H^2(a)}\right)
\end{align}
In  an {\it N}-body  simulation  of DGP  this equation  is  solved at  every
time-step to determine the fifth-force $\frac{1}{2}\nabla \phi$ which is needed to propagate the particles using the geodesics equation
\begin{align}
\ddot{{\bf x}} + 2H{\bf \dot{x}} = -\frac{\nabla \Phi}{a^2} = -\frac{\nabla \Phi_N}{a^2} -\frac{\nabla \phi}{2a^2}
\end{align}
The original DGP  model, which has self-accelerating cosmological solutions, is ruled out by observations and by problem of the ghost in the gravitational sector \citep{lrr-2010-5}. The model  we study here is  the so-called
normal-branch  DGP model  where the  acceleration of  the Universe  is
driven by a cosmological constant  just as in $\Lambda$CDM. This model
is a useful toy-model to study the particular screening mechanism, the
so-called Vainshtein mechanism \citep{1972PhLB...39..393V}, used by  DGP to hide the modifications
of gravity in  local experiments. The modifications of  gravity in the
vicinity of  a massive object  of mass $M$  are determined by  a scale
known as  the Vainshtein  radius which  for DGP is  given by  $r_{V} =
\left(\frac{16   r_c^2   GM}{9\beta^2}\right)^{1/3}$.   Test-particles
outside  the  Vainshtein radius  will  feel  a modified  gravitational
force, $F_{\rm  eff} =  F_N\left(1+\frac{1}{3\beta(a)}\right)$, while
test-particles far  inside the  Vainshtein radius  will just  feel the
standard Newtonian gravitational force.

\subsection{ The symmetron model}

The symmetron model was originally proposed in \cite{hinter} (see also \cite{2008PhRvD..77d3524O,2005PhRvD..72d3535P}). The action of the symmetron model is given by
\be
S = \int \sqrt{-g}{\rm d}^4x \left[ {\frac{M_{\rm Pl}^2}{2}} R - \frac{1}{2}\nabla_a\phi\nabla^a\phi - V(\phi)\right] + S_M(\tilde{g}_{ab}, \psi) \ ,
\label{symm-action}
\ee
where \scb{$R$ is the Ricci scalar}, the Einstein and Jordan frame metrics ($g_{ab}$ and $\tilde{g}_{ab}$) are conformally related
\be
\tilde{g}_{ab} = A^2(\phi) g_{ab},
\ee
\scb{and $S_M$ is the matter action which describes the evolution of the matter fields $\psi$.}
The potential and conformal factor that define the model are
\begin{eqnarray}
V(\phi) &=& -\frac{1}{2}\mu^2\phi^2 + \frac{1}{4}\lambda\phi^4 + V_0 \\
A(\phi) &=& 1 + \frac{1}{2}\left(\frac{\phi}{M}\right)^2,
\end{eqnarray}
where $\mu$ and $M$ are mass scales, $\lambda$ is a dimensionless constant, and $V_0$ is set to match the observed cosmological constant. The equation of motion for the scalar field that comes out from the action (assuming non-relativistic matter) is
\be
\square\phi = \frac{dV(\phi)}{d\phi} + \frac{dA(\phi)}{d\phi}\rho,
\label{eq_motion_phi}
\ee
By fixing the metric to be a perturbed Friedmann-Robertson-Walker metric in the Newtonian gauge
\be
\label{metric}
{\rm d}s^2 = -(1+2\Phi) dt^2 + a^2(1-2\Phi)({\rm d}x^2+{\rm d}y^2+{\rm d}z^2),
\ee
\scb{where $\Phi$ is a scalar perturbation (i.e. the gravitational potential in a classical context)}, we can write the equation of motion of the scalar field in the  form
\be
\nabla^2\phi = \left(\frac{\rho}{M^2}-\mu^2 \right)\phi + \lambda \phi^3 = \frac{dV_{\rm eff}(\phi)}{d\phi},
\ee
where $\rho$ is the matter density and the effective potential is given by
\be
V_{\rm eff}(\phi) = \frac{1}{2}\left(\frac{\rho}{M^2} - \mu^2\right)\phi^2 + \frac{1}{4}\lambda\phi^4 + V_0.
\label{def_effective_potential}
\ee
Note that we have used the approximation $|A(\phi)-1| \ll 1$ to simplify the equation above. From this equation, it is possible to see that the expectation value of the scalar field vanishes at high matter densities. This sets the conformal factor $A$ to unity and thus decouples the scalar from the matter, producing the screening of the fifth force.

To express the equation of motion in a simple form we define a dimensionless scalar field $\chi \equiv \phi/\phi_0$, where $\phi_0$ is the expectation value for $\rho=0$:
\be
\phi_0 = \frac{\mu}{\sqrt{\lambda}}.
\ee
We also substitute the three free parameters $(M, \mu, \lambda)$ and use instead the range of the field that corresponds to $\rho=0$,
\be
\lambda_0 = \frac{1}{\sqrt{2}\mu} \ ,
\label{def_lambda0}
\ee
a dimensionless coupling constant,
\be
\beta_s = \frac{\phi_0 M_{\rm Pl}}{M^2} \ ,
\label{def_beta}
\ee
and the scale factor at the time of symmetry breaking,
\be
a_{\rm SSB}^3 = \frac{\rho_0}{\rho_{\rm SSB}} = \frac{\rho_0}{\mu^2 M^2} \ ,
\label{def_assb}
\ee
where $\rho_0 = 3\Omega_m H_0^2 M_{\rm Pl}^2$ is the background density at $z=0$. We also define the associated redshift $z_{\rm SSB} = 1/a_{\rm SSB} - 1$ which is the redshift for which the modifications of gravity start to kick in cosmologically. With these variables the equation for the dimensionless scalar field $\chi$ is then
\be
\nabla^2\chi = \frac{a^2}{2\lambda_0^2}\left[\left(\frac{\rho}{\rho_{\rm SSB}} - 1\right)\chi + \chi^3 \right].
\label{eq_motion_chi}
\ee
The effects of the scalar field on the matter distribution in a cosmological {\it N}-body simulation will be given by a modification of the geodesics equation, which takes the following form:
\be
\label{geo_code}
\ddot{\mathbf{x}} + 2 H \dot{\mathbf{x}} +
    \frac{\nabla\Phi}{a^2} +
     \frac{6\Omega_m H_0^2}{a^2} \frac{(\beta_s\lambda_0)^2}{a_{\rm SSB}^3} \chi\nabla\chi = 0.
\ee
\scb{Here $H_0$ is the Hubble parameter at redshift $z=0$, $\Omega_m$ is the mean matter density at redshift $z=0$ normalised to the critical density,} and the dots represent derivatives with respect to Newtonian time defined by equation~\ref{metric}.

\subsection{The $f(R)$ model}

Among the large number of $f(R)$ models that exist in the literature we choose the well known Hu-Sawicky model presented in \cite{hu}. The action that defines the model is
\be
S = \int \sqrt{-g}{\rm d}^4x \left[ \frac{R+f(R)}{16\pi G} + L_M \right],
\ee
where the free function $f$ is chosen as
\be
f(R) = - m^2\frac{c_1(R/m^2)^n}{c_2(R/m^2)^n+1},
\ee
where $m^2 \equiv H_0^2\Omega_m$ and $c_1$, $c_2$, and $n$ are dimensionless model parameters. By requiring the model to give dark energy, it is possible to reduce the number of free parameters from three to two ($n$ and $f_{R0}$).  This requirement translates into
\be
\frac{c_1}{c_2} = \frac{6\Omega_\Lambda}{\Omega_m},
\ee
\scb{where  $\Omega_{\Lambda}$ is the density parameter associated with the cosmological constant}.
Instead of using $c_1$ (or $c_2$) as the second free parameter, it is convenient to use
\be
f_{R0} = -n\frac{c_1}{c_2^2}\left(\frac{\Omega_m}{3(\Omega_m+4\Omega_\Lambda)}\right)^{n+1},
\ee
which relates to the range of fifth force in the cosmological background at redshift $z=0$ as
\be
\lambda_0 = 3 \sqrt{\frac{(n+1)}{\Omega_m+4\Omega_\Lambda}}\sqrt{\frac{|f_{R0}|}{10^{-6}}}~~~ \mbox{Mpc}/h,
\ee
where $\lambda_0$ is the range of the field, which is typically given in Mpc$/h$. General Relativity is formally recovered in the limit $f_{R0}\to 0$.

In the quasi-static limit \citep{2015JCAP...02..034B}, the scalar field $f_R$ fulfils the following equation of motion,
\begin{eqnarray}
\nabla^2f_R = &-\frac{1}{a}\Omega_m H_0^2\delta  + a^2\Omega_m H_0^2 \times \nonumber\\
&\times \left[\left(1+4\frac{\Omega_\Lambda}{\Omega_m}\right)\left(\frac{f_{R0}}{f_R}\right)^{\frac{1}{n+1}}  - \left(a^{-3} + 4\frac{\Omega_\Lambda}{\Omega_m}\right)\right],
\end{eqnarray}
where $f_{R0} = f(R_0)$, $R_0$ is the present value of the Ricci scalar in the cosmological background and $\delta$ is the local matter overdensity in units of the mean density of the Universe.

The geodesic equation takes the  form
\begin{equation}
\ddot{\bf{x}} + 2H \dot{\bf{x}} + \frac{\nabla\Phi}{a^2} - \frac{1}{2}\frac{\nabla f_R}{a^2}= 0,
\end{equation}
where the last term corresponds to the fifth force.

\subsection{Quintessence models}

Quintessence coupled dark energy (\cde) has been proposed by several authors \citep[e.g.][]{Wetterich:1995, Zlatev:1999, Amendola:2000} as an alternative to the standard \LCDM,
in an attempt to solve the so called fine tuning and coincidence problems of the model.
These theories feature a scalar field $\phi$\footnote{ Here and above $\phi = \phi(a)$ denotes the cosmological value of the field $\phi$. } which can be non-minimally coupled to the dark matter, effectively acting as a fifth-force on DM particles.
The general Lagrangian for this class of theories reads:

\begin{equation}\label{eq:lagrangian}
L = \int {\rm d}^4x \sqrt{-g} \left(-\frac{1}{2}\partial_{\mu}\partial^{\mu}\phi
	+ V(\phi)+ m(\phi)\psi_{m}\bar{\psi}_{m} \right)
\end{equation}

\noindent
where the matter field $\psi_m$ is allowed to interact with $\phi$ through the $m(\phi)$, the mass term.
A popular choice for the self interaction potential $V(\phi)$ is of the inverse power-law kind \citep{Ratra:1988}:

\begin{equation}\label{eq:ratra}
V(\phi) = V_0\left(\frac{\phi}{M_p}\right)^{-\alpha}
\end{equation}

\noindent
while the mass-mediated interaction term is chosen

\begin{equation}\label{eq:mass}
m(\phi) = m_0 \exp{\left(\beta(\phi)\frac{\phi}{M_p}\right)}
\end{equation}

In the following analysis, we will consider a simple realisation of this coupled dark energy model (\cde) with a constant $\beta(\phi) = \beta_0$ and
a positive $\alpha$ for the potential.


\section{Simulations}\label{sec:simulations}
\begin{table}
\begin{center}
\caption{Model parameters for the symmetron, $f(R)$, DGP and \cde\ runs.
The range of the field in the $f(R)$ model, $\lambda_0$, is derived from the value of $f_{R0}$ and is given in \hMpc.
\label{tab:model_parameters}
}
  \begin{tabular}{lccc}
 \hline
    Model &$\beta_s$&$z_{\rm SSB}$&$\lambda_0$ (\hMpc)\\
 \hline
    \symma & 1 & 1 & 1\\
    \symmb & 1 & 2 & 1\\
 \hline
    Model &$n$ & $|f_{R0}|$ & $\lambda_0$ (\hMpc)\\
 \hline
    \frfou & 1 & $10^{-4}$& 23.7\\
    \frfiv & 1 & $10^{-5}$& 7.5\\
    \frsix & 1 & $10^{-6}$& 2.4\\
 \hline
    Model & $r_cH_0$ & $r_c$ (\hGpc)  &   \\
 \hline
    \dgpot  & 1.2 & 3.6   &  \\
     \dgpfs & 5.6 & 16.8 &  \\
\hline
	Model & $\alpha$ & $\beta_0$ \\
     \hline
	\cde & $-0.137$ & 0.099 \\
 \hline
  \end{tabular}

\end{center}
\end{table}

\begin{table}
\begin{center}
\caption{Simulation settings for the various models. Box-sizes are expressed in units of \hMpc.}
\label{tab:models}\begin{tabular}{lccc}
\hline
Model & Box & $N_{\rm part}$ \\
\hline
\LCDM-I & 250 & $512^3$ \\
\frfou\ & 250 & $512^3$ \\
\dgpot\ & 250 & $512^3$ \\
\dgpfs\ & 250 & $512^3$ \\
\hline
\LCDM-II & 256 & $512^3$ \\
\frfiv\ & 256 & $512^3$ \\
\frsix\ & 256 & $512^3$ \\
\symma\ & 256 & $512^3$ \\
\symmb\ & 256 & $512^3$ \\
\hline
\LCDM-III & 250 & $2\times1024^3$ \\
\cde\ & 250 & $2\times1024^3$ \\
\hline
\end{tabular}
\end{center}
\end{table}

\begin{table}
\begin{center}
\caption{Cosmological parameters used for the different realisations of \LCDM.}
\label{tab:params}\begin{tabular}{lcccc}
\hline
Model & $\Omega_{\Lambda}$ & $\Omega_M$ & $h$ & $\sigma_8$\\
\hline
\LCDM-I & 0.733 & 0.267 & 0.719 & 0.8 \\
\LCDM-II & 0.65 & 0.35 & 0.65 & 0.8 \\
\LCDM-III & 0.73 & 0.27 & 0.70 & 0.8 \\
\hline
\end{tabular}
\end{center}
\end{table}

Simulating a fifth-force kind of interaction requires modifying standard {\it N}-body solvers. Here we will briefly resume the main ideas and properties of such codes.
We considered a total of three simulation series, each one of which has been ran with the same initial random seed and
a realization of the \LCDM\ cosmology, which is used as a benchmark.
Halo catalogues have been extracted using the \texttt{AHF} halo finder \citep{Knollmann:2009}.
In all the \LCDM\ and \cde\ simulations, the power spectrum of the initial conditions were
normalised using the redshift zero $\sigma_8$. For $f(R)$, symmetron and DGP models the $\sigma_8$ was normalized to be the same 
at the \emph{starting} redshift, leading to slightly higher $z=0$ results. However, these small changes are not expected to play
any substantial role at the sub-megaparsec scales considered in this analysis.

\subsection{DGP, symmetron and f(R)}
The simulations were run with the code \texttt{ISIS} \citep{isis} which is a modified gravity modification of \texttt{RAMSES} \citep{2002A&A...385..337T}.  The code is a particle mesh code which includes adaptive mesh refinements.  In order to solve the equations for the scalar field, the code uses a non-linear version of the linear multigrid solver in \texttt{RAMSES}.  The solver works by doing Gauss-Seidel iterations on the discretised version of the equations to find improved solutions based on an initial guess.  Given the multiscale properties of the problem, the solver also uses the multigrid method to increase the speed of convergence.
In these simulations we used a coarse-level grid with $512^3$ grid cells and each cell was refined if the number of particles
contained in it exceeded $8$. The maximum refinement-level obtained in the simulations was $6$ corresponding to a smallest gridcell of size $7.6-7.8$ \hkpc.
Table \ref{tab:model_parameters} summarises the model parameters for the modified gravity theories. All the simulations were run using the same initial conditions. This is valid since at early times $z\lesssim 2$ the modifications of gravity, in all of the models simulated, have very little impact on the growth of structures.

To generate the only set of initial conditions we used the package \texttt{COSMICS} \citep{Bertschinger:1999}.
Two box sizes have been used for these simulations, of $256$ and $250$\hMpc, while the number of DM particles is $512^3$.

The background cosmology is also the same for all the simulations and is defined as a flat \LCDM,
for which two realizations (one for each box size) have been run using the parameters shown in \Tab{tab:params}.
All the simulations have the same normalisation.
The simulations were run up to redshift zero. Furthermore, all the simulations use the same background cosmology with exactly the same initial conditions.
The samples used for the analysis include all the halos reported by the halo finder with no discrimination between virialized and non-virialized objects.
The halo catalogue has a cut-off for low-mass halos at 20 particles per halo, which corresponds to a minimum halo mass of $1.85\times10^{11}$\hMsun.

\subsection{Quintessence}

The code used to simulate coupled quintessence was described in \citet{Carlesi:2014a},
and implements the algorithm of \citet{Baldi:2010a} on the publicly available code \texttt{GADGET2} \citep{Springel:2005}.

The values for $\alpha$ and $\beta_0$ shown in \Tab{tab:model_parameters} 
are chosen in order to be in agreement with WMAP7 \citep{Pettorino:2012} constraints.
The simulations were ran in a $250$\hMpc\ side periodic box using
$2\times1024^3$ both baryonic and DM particles, with a softening length of $8$ \hkpc\ for DM and baryonic particles. 
The adiabatic approximation was used for the baryonic SPH solver.
As for the previous cases, along with \cde\ we also simulate a standard \LCDM\ cosmology set up with the identical random phase for
the generation of the initial condition, which enables us to consistently cross-correlate objects among the different models.

The algorithm used in the modified code is based on the standard \emph{Tree-PM}, modified in order to
take into account long-range interactions mediated by the scalar field, which affect the DM particles only.
This interaction turns out to act effectively as a rescaling of the gravitational constant, which can be written as
$$ G^{\rm eff}_{\rm DM} = G_N (1 + 2\beta^2(\phi)),$$ where $G_N$ takes the standard Newtonian value.
Moreover, we need to take into account the effect of cosmic friction, which is an additional
quintessence mediated force proportional to $\beta(\phi)\overrightarrow{v}$.
The factors above require to compute the solution of the
Green functions separately for each kind of particle (whether baryonic or DM), as the
additional dark energy interaction may or may not be present.
\noindent
The code uses a set of pre-computed tables of quantities such as the Hubble function $H(a)$,
which are then read and interpolated at run time. This saves computational time, sparing the need to solve
complex systems of equations on the fly at each time step.
Initial conditions have been generated using a suitably modified version of the \texttt{N-GenIC} code \footnote{http://wwwmpa.mpa-garching.mpg.de/gadget/right.html\#ICcode}.


\section{Methods}\label{sec:methods}

The analysis presented here relies on the concept of the \emph{Local Group model}, 
formalized by \citet{Carlesi:2016b} in the context of LG constrained simulations \citep{Carlesi:2016a}.
In this approach, the properties that are used to select
LG-like objects in cosmological simulations are explicitly treated as Bayesian priors, 
expressing our previous knowledge and our prejudices on the system at hand.
In principle, the number of variables that can be used in the definition of the LG and its members is potentially infinite: besides mass, position and velocity, other properties can be employed, 
such as 
stellar mass \citep{Guo:2015}, dwarf galaxies \citep{Busha:2011, Boylan-Kolchin:2013},
Hubble flow \citep{Karachentsev:2009}, filamentary environment \citep{Libeskind:2015a, Carlesi:2016a}.
Therefore, any LG definition of is to some extent arbitrary. This is why we emphasize here
the role played by our choice of the variables used to define it.
Such choices need to be flexible enough to build statistically significant samples of objects
that, by some metric, are akin to the real LG.
These samples can be used to produce the PDFs of LG-related variables and their combinations.
In Bayesian terms, these are conditional probabilities, i.e. functions that express our expectation about a given variable
assuming a specific prior model for both cosmology and the LG.

\subsection{The Local Group model}
The use of a LG model allows us to interchange cosmological models and LG definitions, in order to highlight the role played by
cosmology in shaping the expected properties of LG-like objects.
In this way, we estimate how frequently within a given theory we expect to observe 
\emph{actual} values of LG-variables such as the radial velocity (\vrad) between MW and M31.
Moreover, consistency with the actual LG can be analysed from the viewpoint of quantities such as energy and
angular momentum, which due to the isolation of the system are thought to be almost exactly conserved \citep{Forero-Romero:2013}.
This amounts at determining whether a given cosmology may give rise (and at what rate) to perturbations that
can later evolve into an LG-kind of object. This approach and the results derived from it
are discussed in \Sec{sec:semiconserved}.
We stress again that computing the above quantities we treat haloes as point-like particles. Therefore, concerns about
the limited resolution of the simulations are secondary here, as the internal structure of the haloes plays a substantially negligible
role, as discussed in \App{appendix} in more detail.
This has been tested using \LCDM\ simulations with the same box size but with different numbers of particles.
In fact, it can be shown that both the total number of LG-like pairs and the distribution functions of their properties, 
within the mass ranges which are relevant for the present study, are not affected nor biased by the resolution.

\begin{table}
\begin{center}
\caption{Kinematic priors on velocities (\vtan\ and \vrad, in \kms), relative distances $r$ of the haloes (in \hMpc) and masses (in $10^{12}$\hMsun\ units).
The first set \modz\ is very broad and can be used to derive PDFs for all of the variables, assuming almost no prior knowledge of the mass, separation
and \vrad. In \modo\ we include LG-like objects with negative \vrad\ only, also restricting $r$ and \mlg\ values.
\modtw\ and \modth\ define objects whose dynamics is within $\pm 2\sigma$ from the values of $r$, \vrad\ and \mlg\ of \citet{Marel:2012};
while using $\pm 1\sigma$ intervals around the \vtan\ values of \citet{Sohn:2012} (\vtanI) and \citet{Salomon:2016} (\vtanII).
}
\label{tab:priors}
\begin{tabular}{ccccc}
\hline
\quad & \vrad & \vtan & $r$ & \mlg\ \\
\hline
\modz  & $[-500, 500]$ 	& $[0, 500]$  	& $[0.25, 1.50]$ & $[1, 10]$\\
\modo  & $[-500, 0]$	& $[0, 500]$	& $[0.25, 0.78]$ & $[1, 5]$\\
\modtw & $[-125, -95]$ 	& $[0, 34]$  	& $[0.44, 0.60]$ & $[1, 5]$\\
\modth & $[-125, -95]$ 	& $[100, 225]$  & $[0.44, 0.60]$ & $[1, 5]$\\
\hline
\end{tabular}
\end{center}
\end{table}

\begin{table}
\begin{center}
\caption{Number of selected LG-like pairs per simulation.
$N_{\rm \modz}$, $N_{\rm \modo}$, $N_{\rm \modtw}$ and $N_{\rm \modth}$ correspond to the sample size of pairs that satisfy the
kinematic priors shown in \Tab{tab:models}.
While within some cosmologies \modtw\ LGs are found at a rate comparable with \LCDM,
\symma, \symmb, \frfou\ and \frfiv\ are largely incapable of accounting for that kind of dynamics.
\label{tab:lg_n}
}
\begin{tabular}{ccccc}
\hline
Model & $N_{\rm \modz}$ & $N_{\rm \modo}$ & $N_{\rm \modtw}$ &  $N_{\rm \modth}$ \\
\hline
\LCDM -I  & 7041 & 1452 & 15 & 18 \\
\frfou  & 6770 & 1373 & 3 & 7 \\
\symma  & 8290 & 1656 & 3 & 20 \\
\symmb  & 8168 & 1620 & 0 & 2 \\
\hline
\LCDM -II  & 8929 & 1858 & 19 & 33 \\
\frfiv  & 9278 & 1792 & 6  & 35 \\
\frsix  & 9877 & 1969 & 12 & 32 \\
\dgpot  & 8827 & 1768 & 19 & 34 \\
\dgpfs  & 8827 & 1875 & 16 & 38 \\
\hline
\LCDM -III  & 7633 & 1738 & 15 & 30 \\
\cde\  & 7143 & 1554 & 10 & 29 \\
\hline
\end{tabular}
\end{center}
\end{table}

\subsection{Implementation}
We start defining a LG-like object as a pair of isolated haloes. This requirement is motivated by the fact that the mass budget
of the real LG is dominated by the total mass of MW and M31.
Isolation is defined as the absence of a third object of mass larger or equal than the one of the smallest
halo of the pair within a radius of $2.5$\hMpc\ from the centre of mass of the system.
On top of these two general criteria, a series of priors on the velocities, masses and separations
among these objects are imposed, gradually restricting the range of variation of such parameters to enforce
a stricter resemblance to the real observed system.
\Tab{tab:priors} shows the four ranges of these priors, which define our LG models.
\modz\ is a very general model, where broad criteria are imposed to define LGs from the global number of isolated pairs.
This sample is useful to study \emph{all} the kinematic variables of the system, assuming a very superficial knowledge of the same.
In other words, it can be used to answer the question: \emph{what kind of dynamics do we expect from a pair of close-by, isolated haloes,
within a given cosmological framework?}. \\

\noindent
On the other hand, a more realistic LG model needs to reflect some more important facts about the nature of the M31-MW pair.
This is done within \modo, which implements a more detailed knowledge of the system into the priors.
In this model \vrad\ is constrained to negative values, the range of values for \mlg\ and $r$ are reduced, while keeping the number of object large enough to be statistically significant, as shown in \Tab{tab:lg_n}.
Such a definition overlaps with to the ones used e.g. by \citet{Forero-Romero:2013, Gonzalez:2014, Sawala:2014, Libeskind:2015a} and \citet{Carlesi:2016a}.\\

\noindent
The last two models, \modtw\ and \modth, identify \emph{realistic} LGs, i.e.
objects whose mass, velocity and separation values fall within 2$\sigma$
from the observational data \citep{Marel:2012}. 
Each model implements one of the conflicting measures existing for the tangential velocity of M31:
a low-\vtan\ one, taken from \citet{Sohn:2012} and referred to as \vtanI, and a high-\vtan\ obtained by \citet{Salomon:2016} (\vtanII\ hereafter).
In the following sections, we will take a closer look at the kinematics and dynamics of the LG using samples
drawn from each simulation using these models.

\section{LG dynamics}\label{sec:results}

\begin{table*}
\begin{center}
\caption{Peak likelihood values for masses ($\log_{10}$ in \hMsun\ units) and velocities (\kms) for each cosmology using different LG models, together
with their $95\%$ confidence intervals.
Distributions relative to the \modtw\ and \modth\ samples are not shown due to the smallness of the sample size in both cases.}
\label{tab:m_v_lg}
\begin{tabular}{c}
\begin{tabular}{ccccccc}
\hline
\multicolumn{4}{c}{\qquad\quad\qquad\qquad\modz} & \multicolumn{3}{c}{\modo} \\
\hline
Model & \mlg & \vrad & \vtan & \mlg & \vrad & \vtan \\
\hline
\LCDM -I & $12.44\pm0.30$ & $-101^{+40}_{-54}$ & $72^{+48}_{-30}$ &
$12.35\pm0.20$ & $-124^{+36}_{-40}$ & $78^{+44}_{-33}$ \\
\frfou\ & $12.45\pm0.30$ & $-118^{+49}_{-66}$ & $92^{+63}_{-41}$ &
$12.35\pm0.20$ & $-150^{+51}_{-70}$ & $98^{+55}_{-41}$ \\
\symma\ & $12.46\pm0.29$ & $-114^{+45}_{-58}$ & $74^{+52}_{-32}$ &
$12.35\pm0.21$ & $-153^{+45}_{-44}$ & $82^{+54}_{-33}$ \\
\symmb\ & $12.45\pm0.31$ & $-144^{+59}_{-78}$ & $95^{+65}_{-42}$ &
$12.35\pm0.22$ & $-198^{+69}_{-65}$ & $113^{+63}_{-49}$ \\
\hline
\LCDM -II & $12.36\pm0.32$ & $-90^{+39}_{-55}$ & $69^{+46}_{-29}$ &
$12.30\pm0.22$ & $-121^{+40}_{-42}$ & $79^{+43}_{-33}$ \\
\frfiv\ & $12.36\pm0.32$ & $-106^{+46}_{-64}$ & $82^{+58}_{-46}$ &
$12.32\pm0.23$ & $-145^{+47}_{-55}$ & $95^{+54}_{-39}$ \\
\frsix\ & $12.38\pm0.33$ & $-103^{+44}_{-60}$ & $75^{+52}_{-32}$ &
$12.32\pm0.23$ & $-141^{+43}_{-49}$ & $89^{+51}_{-39}$ \\
\dgpot\ & $12.36\pm0.33$ & $-92^{+40}_{-58}$ & $73^{+51}_{-31}$ &
$12.30\pm0.22$ & $-123^{+40}_{-43}$ & $78^{+47}_{-33}$ \\
\dgpfs\ & $12.36\pm0.32$ & $-91^{+41}_{-55}$ & $69^{+48}_{-28}$ &
$12.29\pm0.22$ & $-120^{+38}_{-45}$ & $79^{+46}_{-34}$ \\
\hline
\LCDM -III & $12.46\pm0.26$ & $-96^{+44}_{-57}$ & $70^{+53}_{-34}$ &
$12.38\pm0.17$ & $-120^{+39}_{-44}$ & $77^{+50}_{-37}$ \\
\cde\ & $12.46\pm0.28$ & $-112^{+48}_{-59}$ & $82^{+56}_{-36}$ &
$12.38\pm0.17$ & $-149^{+43}_{-42}$ & $89^{+54}_{-37}$ \\
\hline
\end{tabular}

\end{tabular}

\end{center}
\end{table*}

We will now study three aspects of the LG dynamics in order to present a comprehensive picture of the possible observational signatures that characterize the models under analysis.
First, we will look at compatibility with observational data, counting the 
number of halo pairs whose properties fall within the allowed confidence intervals.
This enables us to evaluate and compare the expected rate of formation of
LGs in a non standard model and in \LCDM, in a very straightforward way. Second, using a more general LG model yielding larger samples, we compute distribution functions for masses and velocities. These PDFs are then used to compute the average expected dynamics within each model, establishing a link between this cosmology and properties on astrophysical scales. As a last step, we will look at the semi-conserved quantities of the
system, energy and angular momentum, to reduce the influence of transient factors that could affect the previous results.

\subsection{Realistic local groups}

We define as realistic LGs those halo pairs whose values of $r$, \vrad\ and \mlg\ fall within 2$\sigma$ from the values of \citet{Marel:2012}. On top of these, we use two different  1$\sigma$ priors for \vtan : \modtw\ implements the \vtanI\ measurement of \citet{Sohn:2012} while \modth\ employs the \vtanII\ value of \citet{Salomon:2016}. 
The objects obtained in this way provide the most accurate representation of a LG in a simulation.
However, the narrow interval of values due to such strict definitions does not allow to gather statistically meaningful halo samples. 
Therefore, to obtain an estimate of the viability of a theory 
we will simply refer to the number of objects complying with these two prior models.\\

In the last two columns of \Tab{tab:lg_n}, it is shown how the \modtw\ and \modth\ sample sizes
are affected by a change of the cosmology.
In the cases of \cde, \frsix\, \dgpot\ and \dgpfs\ we see that (for both \modtw\ and \modth) these numbers do not substantially change in comparison to the benchmark \LCDM-I, \LCDM-II and \LCDM-III simulations.
This means that the aforementioned models are able to reproduce object whose dynamics is compatible with the
one observed for the actual LG \emph{at least at the same rate} of \LCDM.
However, the other models show a different behaviour. In particular, it has be be noticed how implementing a low-\vtan\ prior
the number of haloes found within the \frfou, \frfiv, \symma\ and \symmb\ simulations is drastically reduced, 
indicating that this specific kind of dynamics can hardly be accounted for within those
cosmological frameworks. Most notably, the \symmb\ model has no \modtw-complying pairs, in \symma\ and \frfou\ the number is reduced
five-fold and in \frfiv\ three-fold.
When assuming a high-\vtan\ prior, on the other hand, the number of objects reaches \LCDM\ levels in \symma\ and \frfiv\, and is increased in both
\frfou\ and \symmb\, indicating that \vtanI\ yields more constraining power than \vtanII.

\begin{figure*}
\begin{center}
$\begin{array}{cc}
\includegraphics[height=7.0cm]{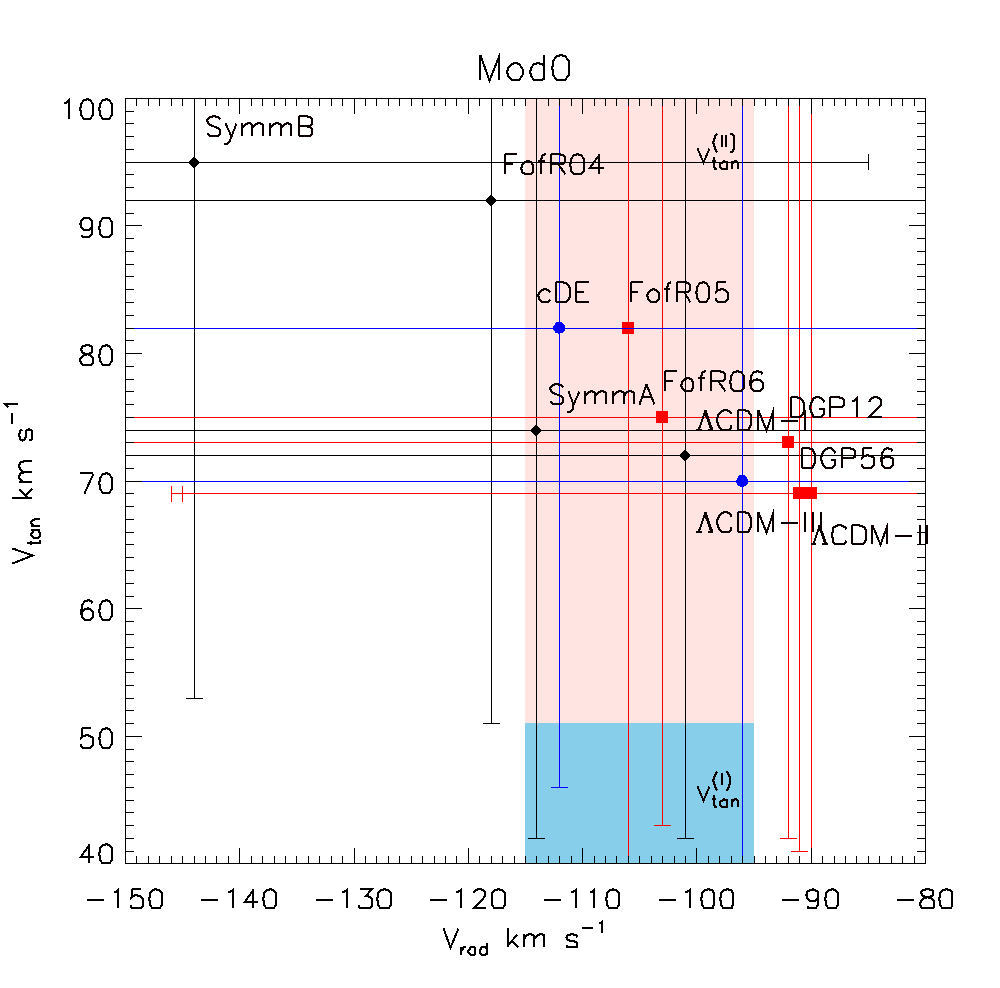} &
\includegraphics[height=7.0cm]{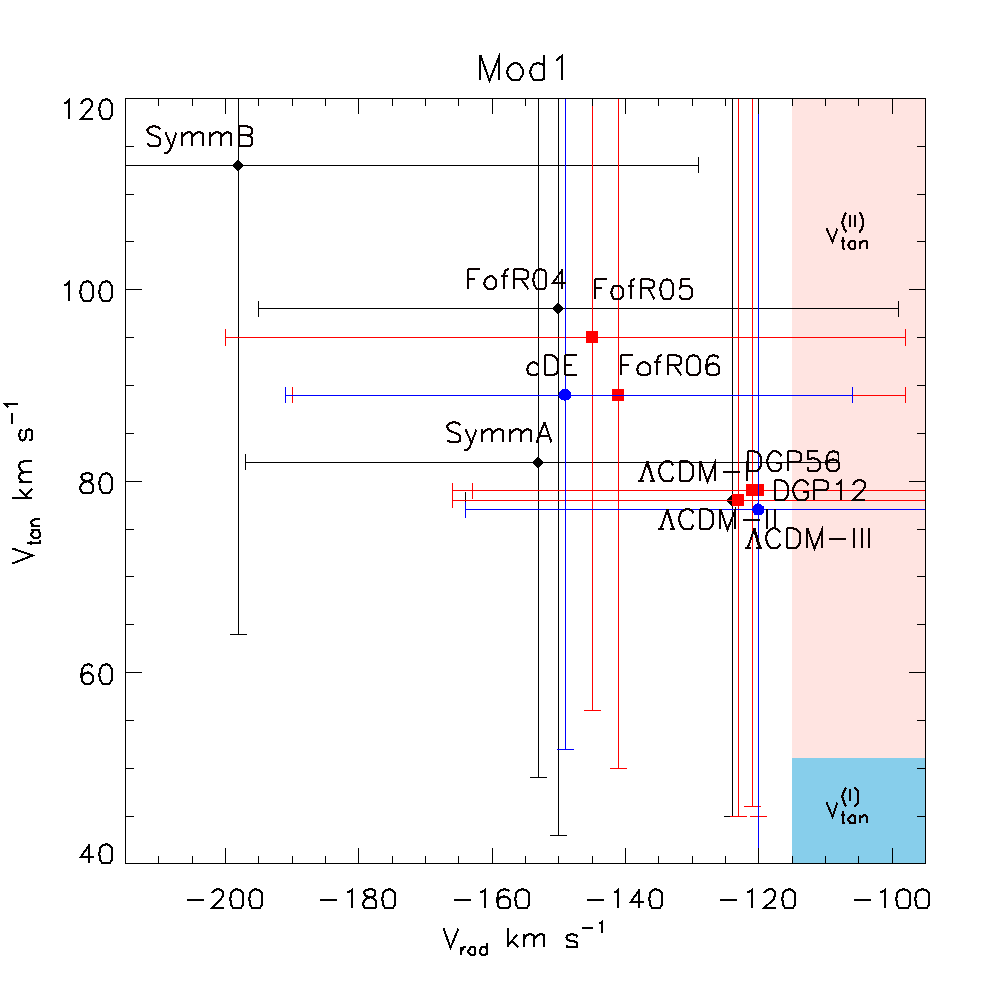}
\end{array}$
\caption{Peak values of \vtan\ versus \vrad\ for the different cosmologies and the two LG models \modz, \modo,
together with their $95\%$ confidence level values.
The shaded regions stretch horizontally for on very narrow interval of values, due to the precision of the \vrad\ measurements.
On the other hand, conflicting \vtan\ measurements lead to a much larger range of $95\%$-allowable values on the $y$ axis.
Models such as \symmb\ can be seen to predict a combination of velocities outside the region allowed by actual measurements.
\label{img:dyn}}
\end{center}
\end{figure*}

\subsection{Kinematic variables}
\noindent
\textbf{Local group masses} Masses for the LG and its two main haloes are not affected by the change in cosmology.
\Tab{tab:m_v_lg} shows that for in each simulation both the peak likelihood and the scatter for the
$\log_{10}$\mlg\ distributions are sensitive to the LG model only and are not affected by modified gravity and \cde.
A small reduction in the mass can be seen on average when switching from \modz\ to \modo, that is,
when selecting pairs with strictly negative \vtan.
Individual masses and the \mmw\ to \mmto\ ratios are also not affected.
We therefore conclude that the mass of the local group and of its members cannot be used as a probe for the kinds of alternative theories considered.
In the case of \cde, this finding is consistent with \citet{Penzo:2015} who found 
a weak dependence of \mvir\ on the cosmology and cam be expected from the negligible
changes to the halo mass function associated with this kind of DE model \citep[see][]{Maccio:2004,Baldi:2010a,Cui:2012,Carlesi:2014a}.
\\

\noindent
\textbf{Radial and tangential velocity}
Factoring the mass parameter out of the following analysis, we seek to establish a direct link between alternative models and expected LG
kinematics. Pairwise velocity has been shown to be strongly affected in modified gravity models by \citet{Hellwing:2014}, so that we expect some correlation between \vrad , \vtan\ and modifications to GR.
In \Fig{img:dyn} we show the peaks of the likelihood in the \vrad - \vtan\ plane for the \modz\ and \modo\ samples,
with error bars and shaded regions indicating the $95\%$ confidence level for models' estimates and observations.
First of all, we remark how the three \LCDM\ simulations largely agree with each other and with the results of \citet{Forero-Romero:2013}
and \citet{Carlesi:2016b}, indicating that the cosmic variance and the different cosmological parameters play here a negligible
role. 
This is true for both \modz\ and \modo\ \LCDM\ samples, so that we can conclude that the reduction
in sample size does not affect our results at this stage.
From both panels we observe that the different cosmologies can predict  several kinds of dynamics
overlapping in some cases only at the two sigma level. However, these differences are enhanced in \modo\, that comprises a
more realistic LG-like kind of objects, i.e. those pairs with negative radial velocity.
In this case we can see how all of the cosmologies, except for DGP which closely mimics \LCDM,
are characterized by peak \vtan\ values substantially larger in module than the observational interval.
In particular, \symmb\ does not overlap with the allowable range of values, meaning that the likelihood
of observing a combination of velocities compatible with the LG would is less than $5\%$.
Moreover, also \symma, \cde, \frfou, \frfiv\ and \frsix\ cannot reproduce the data, preferring \vrad - \vtan\
combinations that only marginally agree with the observations.
In table \Tab{tab:m_v_lg} these values are shown together with their \LCDM\ counterparts: we can see how
the additional interaction affects the peaks, increasing the absolute value of \vtan\ by $\approx60\%$ and \vrad\ by $\approx 50\%$ in \symmb.
These same values $\approx 20-30\%$ larger in the case of the other $f(R)$ and symmetron models.
This enhancement is due to the additional interaction and is proportional to the strength of the coupling,
as can be seen in the case of $f(R)$. We have thus been able to determine that the
fifth-force induced modifications to the expected velocities favour kinematic configurations
largely at odds with the observations.
This result confirms what we had found analysing \modtw\ and \modth\ samples, where the
sharp reduction in the number of viable LGs in the 
\symmb, \frfou\ and \frfiv\ cosmologies also signalled the difficulty of such theories to account 
for the real LG dynamics..

\label{sec:semiconserved}
\begin{table}
\begin{center}
\caption{Values and intervals used to generate MC contours.
\mmw, \mmto\ are expressed in $10^{12}$\hMsun\ units,
inter halo distance $r$ in \hkpc\ while \vrad\ and \vtan\ in \kms.
The intervals on $r$ and \vrad\ correspond to the 2$\sigma$ values of \citet{Marel:2012}, while for
\vtan\ they were chosen in agreement with \citep{Sohn:2012}.}
\label{tab:mc}\begin{tabular}{cc}
\hline
\mmw & (0.5, 2.5) \\
\mmto & (0.5, 2.5) \\
$r$	& (440, 600) \\
\vrad	& (-125, -95) \\
\vtan	& (0, 50) \\
\hline
\end{tabular}
\end{center}
\end{table}

\subsection{Global properties}

\begin{figure*}
\begin{center}
$\begin{array}{ccc}
\includegraphics[height=5.0cm]{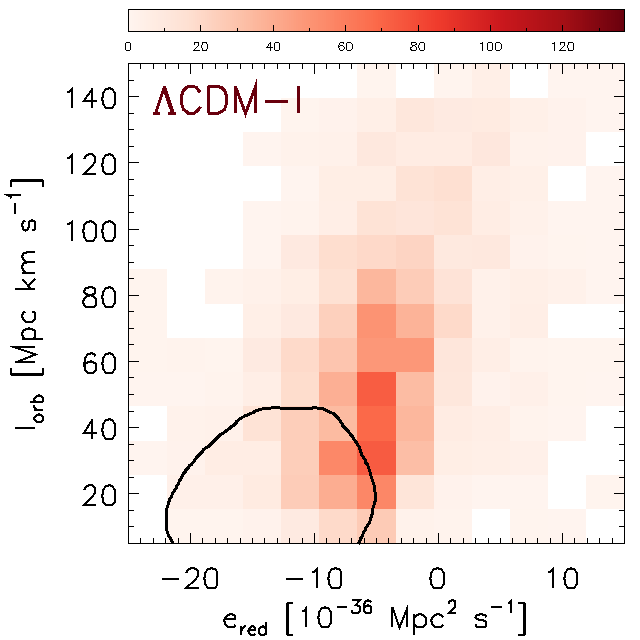} &
\includegraphics[height=5.0cm]{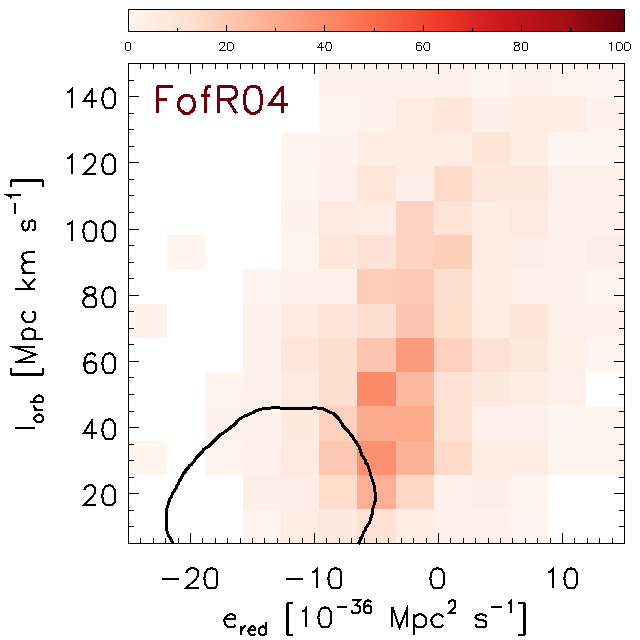} &
\includegraphics[height=5.0cm]{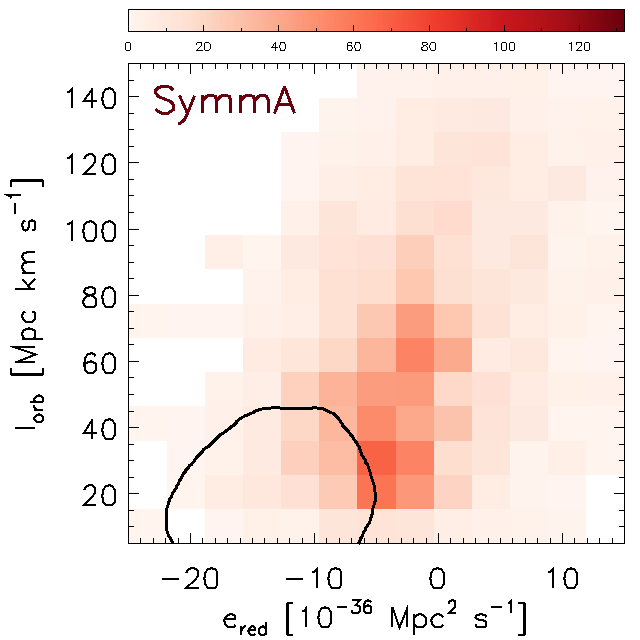} \\
\includegraphics[height=5.0cm]{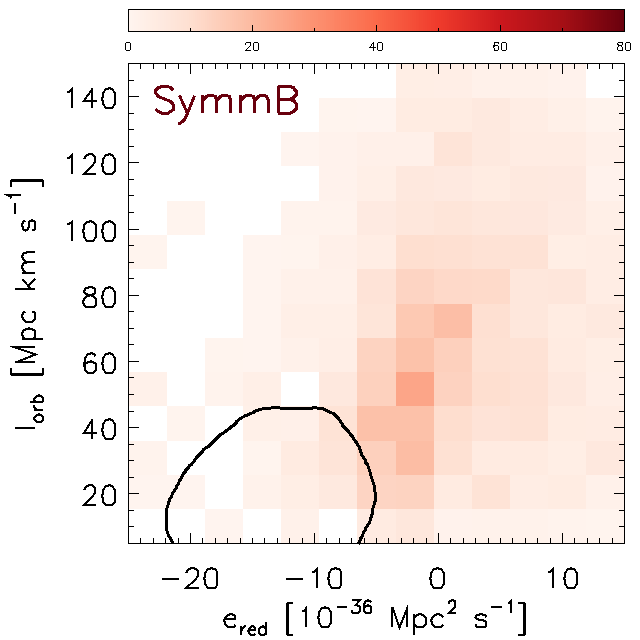} &
\includegraphics[height=5.0cm]{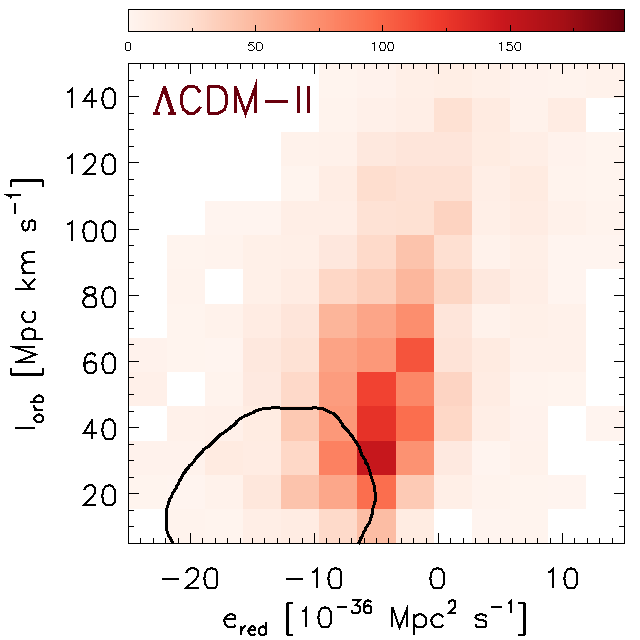} &
\includegraphics[height=5.0cm]{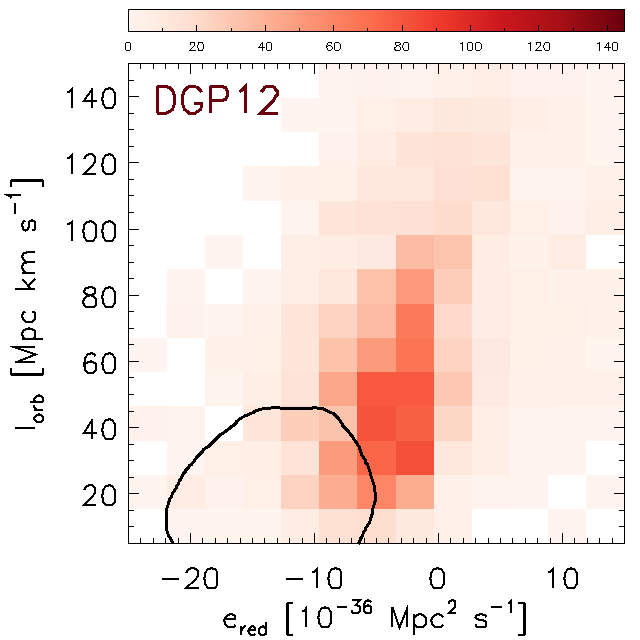} \\
\includegraphics[height=5.0cm]{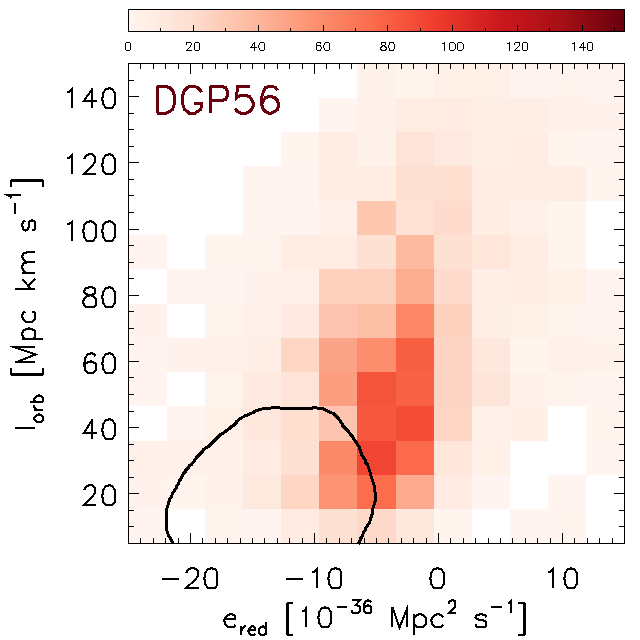} &
\includegraphics[height=5.0cm]{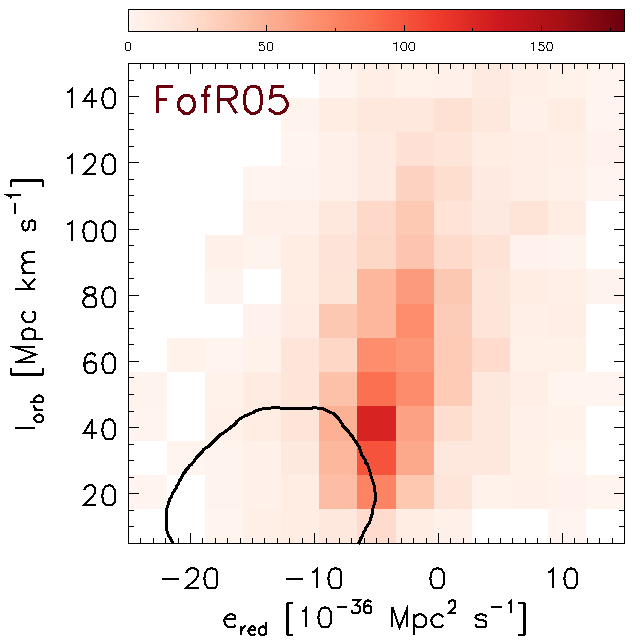} &
\includegraphics[height=5.0cm]{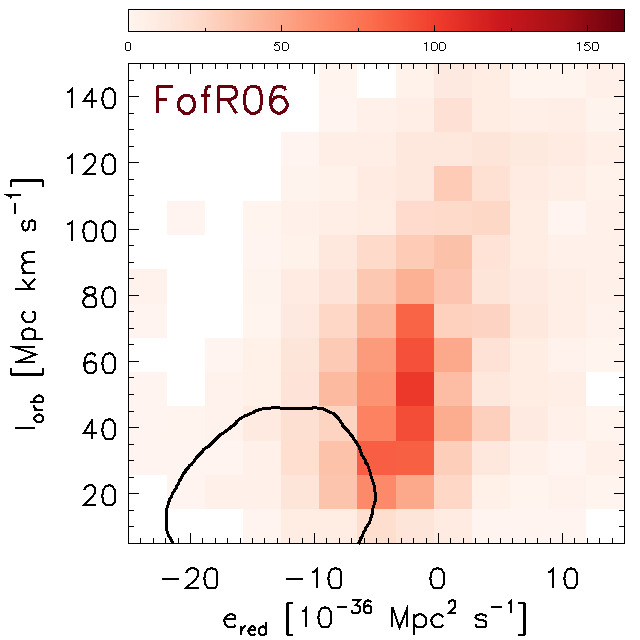} \\

\includegraphics[height=5.0cm]{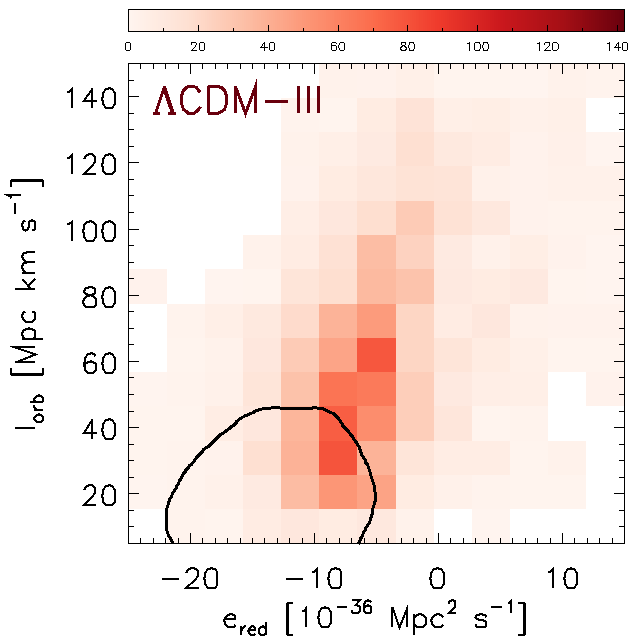} &
\includegraphics[height=5.0cm]{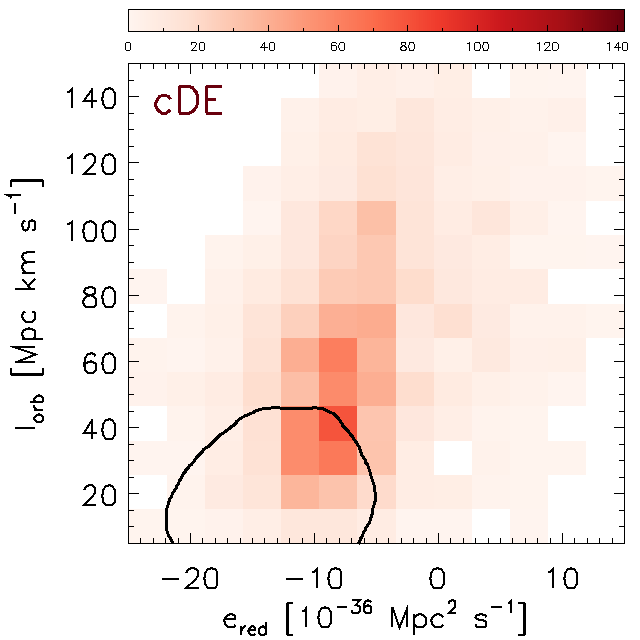} \\
\end{array}$
\caption{LG-like pairs versus MC contours showing the 1$\sigma$ and 2$\sigma$ confidence levels.
Pairs have been chosen and MC intervals have been generated using the low \vtan\ estimate of \citep{Sohn:2012}
\label{fig:mcmw}}
\end{center}
\end{figure*}

\noindent
In addition to the analysis of mass and velocity, which are directly observable and can be straightforwardly compared
to the available data, it is useful to take a look at the global dynamical state of the system, determined by \emph{combinations} of observable variables.
In fact, the capacity of a given cosmological model to produce halo pairs whose properties
are consistent with actual observations of $r$, \vtan\ and \vrad\ at $z=0$
does not rule out the possibility that these transient values might be closer to the real LG ones in
a different moment backwards or forwards in time.
Therefore, a complementary approach consists in analysing combinations of variables, conveying informations about more fundamental properties of the system.\\

\noindent
\textbf{Semi-conserved quantities} We start defining two quantities that characterize the \emph{global} dynamical state of the LG.
Placing ourselves in the MW frame of reference we introduce the reduced total energy:

\begin{equation}\label{eq:ered}
e_{\rm red} = \frac{1}{2}\mathbf{v}_{\rm M31}^2 - \frac{G M_{\rm LG}}{| \mathbf{r}_{\rm M31}|}
\end{equation}

\noindent
and the (reduced) orbital angular momentum:

\begin{equation}\label{eq:l}
l_{\rm orb} = | \mathbf{r}_{\rm M31} \times \mathbf{v}_{\rm M31} |
\end{equation}

\noindent
Ideally, these two quantities would be perfectly conserved if the identified LG-like objects were completely isolated, 
in reality, interactions with smaller nearby haloes spoil their exact conservation.
Moreover, it has to be noticed that \Eq{eq:ered} is derived assuming GR, and therefore is not expected to measure the total (reduced) energy within \cde\ and modified gravity cosmologies. 
However, this variable can still be used in those contexts as an \emph{observationally}-relevant variable, putting aside its original physical meaning. 
In fact, \Eq{eq:ered} is a combination of astrophysical variables that do not rely on cosmology for their measurement. Our aim here is to determine how it is expected to behave 
in modified gravity and coupled dark energy, and compare that to observations in a consistent and model-independent
approach.\\

\begin{table*}
\begin{center}
\caption{Energy, angular momentum, spin parameter and fraction of \modo\ pairs falling within the $95\%$ confidence level of the
MC generated boundaries.\label{tab:semicon}}
\begin{tabular}{ccccc}
\hline
Model & \lorb & \ered & $\log_{10}\lambda$ & $f_{\rm LG}$ \\
\hline
\LCDM-I & $  53.82^{+ 26.06} _{ -22.53}$ & $  -3.96^{   +2.80} _{  -2.89}$ & $  -1.49^{   +0.19} _{  -0.24}$ & 0.16 \\
\frfou & $  59.48^{ +31.69} _{ -24.16}$  & $  -1.65^{   +3.83} _{  -2.99}$ & $  -1.52^{   +0.24} _{  -0.27}$ & 0.08 \\
\symma & $  53.21^{  +29.06} _{ -22.09}$  & $  -2.90^{   +3.71} _{  -3.59}$ & $  -1.54^{   +0.21} _{  -0.28}$ & 0.13 \\
\symmb & $  60.45^{  +29.73} _{ -22.74}$  & $   0.54^{   +5.37} _{  -3.64}$ & $  -1.59^{   +0.27} _{  -0.31}$ & 0.05\\
\hline
\LCDM-II & $  52.30^{ +26.92} _{ -21.10}$ & $  -3.69^{  +2.59} _{  -3.13}$ & $  -1.49^{   +0.19} _{  -0.26}$ & 0.15 \\
\frfiv & $  59.49^{  +28.99} _{ -22.87}$ & $  -1.38^{   +3.35} _{  -2.69}$ & $  -1.50^{   +0.23} _{  -0.30}$ & 0.06 \\
\frsix & $  56.40^{  +26.08} _{ -22.01}$ & $  -2.25^{   +3.13} _{  -2.98}$ & $  -1.52^{   +0.23} _{  -0.28}$ & 0.09 \\
\dgpot & $  53.60^{  +27.50} _{ -20.38}$ & $  -3.34^{   +2.56} _{  -3.14}$ & $  -1.48^{   +0.19} _{  -0.26}$ & 0.14 \\
\dgpfs & $  52.09^{  +26.25} _{ -20.72}$ & $  -3.60^{   +2.46} _{  -3.15}$ & $  -1.49^{   +0.20} _{  -0.26}$ & 0.14 \\
\hline
\LCDM-III & $  52.11^{  +25.41} _{ -18.88}$ & $  -3.41^{  +2.35} _{  -2.47}$ & $  -1.48^{   +0.23} _{  -0.28}$ & 0.19 \\
\cde	 & $  56.43^{  +27.96} _{ -18.63}$ & $  -2.98^{  +2.41} _{  -2.36}$  & $  -1.59^{   +0.24} _{  -0.28}$ & 0.15 \\
\hline
\end{tabular}
\end{center}
\end{table*}

To compare \ered\ and \lorb\ with the data, we follow the procedure of \citet{Forero-Romero:2013}, drawing contours in the \ered -\lorb\ plane to identify
those regions of the parameter space that are compatible with current observations.
The contours are generated through $10^{7}$ Monte Carlo iterations, where at each step the values of velocity, mass and
distance are drawn from a Gaussian distribution within the intervals shown in \Tab{tab:mc}.
The value of
\vrad\ was taken from \citet{Marel:2012}, $r$ from \citet{Marel:2008}, while \mmw\ and \mmto\ are consistent with
\citet{Marel:2012} and \citet{Boylan-Kolchin:2013}.
Due to the large discrepancy existing among the \vtanI\ and \vtanII\
the MC was ran twice using the two different estimates. It turns out that the intervals generated with \vtanII\ are extremely large and possess no constraining
power, so that we do not consider them in the following analysis. \\

In \Fig{fig:mcmw} we bin the \modo\ objects in the \ered\ - \lorb\ plane and compare them to the observational $95\%$ confidence intervals.
The fraction of objects $f_{\rm LG}$ that falls within those boundaries is shown in the last column of \Tab{tab:semicon}.
This quantity allows us to gauge the ability of a model to account for the observed \emph{global} dynamics of the system disregarding its transient state.
We note that the number for \LCDM\ varies between 0.15 and 0.19 -- which is consistent with \citet{Forero-Romero:2013}, who found a 0.08 to 0.12
fraction at the $68\%$ confidence interval.
In the case of modified gravity cosmologies, the results largely confirm the conclusions obtained 
with the previous analyses of velocities. In fact, models such as \frfou, \frfiv, \symmb\ and - to a lesser extent - \frsix\ produce LG-like pairs within the observational boundaries at substantially lower rates than \LCDM. In particular, the fraction of \modo\ LGs falling within the $95\%$ confidence level is reduced by a factor of three in \symmb\ and around a factor of two in \frfou\ and \frfiv, showing from another
perspective that the LG dynamics can only be poorly accounted for within these models.
These discrepancies are explained by the increased velocity of M31, 
affecting the values of \ered\ (through the
kinetic energy term in \Eq{eq:ered} and \lorb, that tend to shift the distribution farther away from
the region allowed by the data. 

\noindent
\textbf{Spin parameter}
Besides these two variables, we conclude our analysis taking a look at the dimensionless spin parameter $\lambda$ \citep{Peebles:1971}, which is defined as

\begin{equation}\label{eq:spin}
\lambda = \frac{\mu^{3/2} l_{\rm orb} \sqrt{e_{\rm red}} }{G M_{\rm LG}^{5/2}}
\end{equation}

\noindent
where $\mu =$ (\mmw \mmto)/\mlg\ and G is the Newtonian constant.
Spin parameters for individual haloes are known to be slightly higher in fifth-force cosmologies, as enhanced velocities 
also lead to an increased rotational support of the haloes \citep{Hellwing:2013,Carlesi:2014a,He:2015}.
This is consistent with our findings about the spin parameter distribution: in \Tab{tab:semicon} we clearly see how the median
$\log _{10}(\lambda)$ values tend to be slightly higher in \cde, $f(R)$ and Symmetron cosmologies.
DGP results, on the other hand, are perfectly aligned with \LCDM\, due to the negligible changes to 
pairwise velocities discussed above.
For consistency, we have also checked that 
\LCDM\ results for this quantities are consistent with the values found by \citet{Forero-Romero:2013}. However, this effect is generally rather small and does not yield much constraining power on the models.


\section{Conclusions}\label{sec:conclusions}

In this work we have studied the dynamics of LG-like objects within different cosmologies.
We evaluated the consistency of the observed kinematics of MW and M31
with alternative models of the Universe, in order to gauge their viability under the assumption
that the LG is not an outlier - -- an idea that lies at the core of the \emph{near-field} cosmology approach.
Four different kinds of alternative cosmologies were taken into account: $f(R)$ \citep{hu}, the symmetron \citep{hinter},
DGP \citep{2000PhLB..485..208D} and coupled quintessence \citep[\cde ,][]{Amendola:2000}. The {\it N}-body 
simulations for each model were ran using different parameters and then 
was compared to a benchmark \LCDM\ simulation with an identical random-phase realization of the initial conditions.
All of these non standard models introduce a fifth-force interaction to explain the observed late-time
acceleration of the Universe, and such a modification of the standard gravitational force is expected to affect the dynamics
at LG scales.
Therefore, we aimed at determining how often one expects to find LG-like objects
in simulations of alternative cosmological frameworks, given a definition (model) of the LG, motivated by astrophysical observations.\\

We identified LGs as pairs of isolated
haloes, using four variables for the identification of such objects:
total mass \mlg, inter-halo separation $r$, radial velocity \vrad\ and tangential velocity \vtan.
From these four quantities, three additional variables where derived, i.e. the (reduced) total energy
\ered, (reduced) orbital angular momentum \lorb\ (both of which are almost exactly conserved and thus
convey time-independent information about the system) and dimensionless spin parameter $\lambda$.
All these quantities treat haloes as point-like particles, 
allowing us to neglect their internal structure and lowering the resolution requirements for the simulations. 

We introduced four different LG models, enforcing different priors on the aforementioned variables. The
samples of objects obtained in this way allowed us to carry a three-level analysis for each cosmology, focusing on

\begin{itemize}
\item[(a)] 2$\sigma$-agreement with $r$, \vrad, \mlg and \vtan
\item[(b)] distribution of \vtan - \vrad
\item[(c)] distribution of \ered\ and \lorb
\end{itemize}

\noindent
Point (a) allowed to determine of many LG-like objects could be found with a specific (transient) dynamical
state in agreement with observations. In this case, and in particular using the low-\vtan\ estimate of \citet{Sohn:2012} (\vtanI),
it was possible to establish that \symma, \symmb, \frfou\ and \frfiv\ can hardly account for the observed properties
of the LG, which cannot be reproduced at all (like in \symmb) or can be reproduced at a rate three to five smaller than in \LCDM\
(as in the case of \symma, \frfou\ and \frfiv). On the other hand, high-\vtan\ estimates (the \vtanII\ of \citet{Salomon:2016})
would be in agreement with most of the models, with the exception of \symmb\ and \frfou. \\

However, to address the problems related to the transitive nature of those
variables as well as the statistical significance of the samples, two additional halo samples were built using the broader selection criteria of \modz\ and \modo\ and
study the variables mentioned in point (b) and (c). 

We first noticed that even though different cosmologies to not change the distribution of masses, they predict a very different kind of kinematics,
in particular for halo pairs with negative radial velocities.
Peak values for \vtan\ and \vrad\ are up to $\approx 60\%$ than in \LCDM\ for the most extreme case (\symmb),
and show an average increase of $\approx 25\%$ for \symma, \frfou, \frfiv\ and \cde.
These deviations from the expected \LCDM\ behaviour are a direct consequence of the fifth-force, which acts as
an additional pull alongside gravity, enhancing the relative velocity between the halos.
This picture is consistent with the results of point (c). In fact, as a consequence of higher average velocities,
LG-like pairs have in general larger (less negative) \ered, \lorb\ as well as $\lambda$ values.
This is more clear in the case \symmb, \frfou\ and \frfiv, where
the large increase in the (reduced) kinetic energy term of \Eq{eq:ered} leads, in the most extreme case, to a positive median \ered\ of $   0.54^{   +5.37} _{  -3.64}$, way 
above the quoted \LCDM\ value of $   -3.96^{   +2.80} _{  -2.89}$.
Binning the number of haloes in the \ered - \lorb\ plane, we could determine that same models lead to a fraction of LGs within the $95\%$ observational confidence interval
which is $(2-3)$ times smaller than \LCDM.
\cde, \frsix\ and \symma\ have similar features (of higher \ered, \lorb\ and $\lambda$ values) though the global state of the
system overlaps consistently more with \LCDM. 
\dgpot\ and \dgpfs\, on the other hand, are indistinguishable from the standard models.\\

These results are consistent with other well-known probes.
For the Hu-Sawicky $f(R)$ model the best constraints today from cluster abundances \citep{2015PhRvD..92d4009C}, and Sunyaev-Zel'dovich clusters \citep{2016arXiv160707863P} and from the matter power-spectrum  \citep{2014JCAP...03..046D}) place the \frfou\ model at odds with the data and also disfavour \frfiv\ \citep[see][for a list of other constraints]{2016arXiv160902937G}. Not many groups have performed explicit analysis for the symmetron, though the enhancement of the power-spectrum and the halo mass-function \citep{2012JCAP...10..002B} also show that the \symmb\ model is in tension with observations. \\

To sum up, we have analysed the small-scale regime of a large set of alternative cosmological theories,
highlighting out the effects of an additional fifth-force interaction on the dynamics of the LG. Such an approach has the advantage of not requiring high-resolution simulations to deliver predictions on sub-megaparsec scales.
Applying this kind of analysis to cosmological simulations, we have been able to signal a large difference between the theoretical predictions of some models and
the observations. In those cases, we found that the additional pull is likely to lead to extremely large
relative velocities between the main haloes of the LG, increasing its energy and angular momentum budget.
We have shown that this method is capable of helping in the process of model selection using astrophysical-scale data, which, 
used in combination with other cosmological probes, can lead
to a deeper understanding of the still unsolved mysteries of our Universe.

\section*{Acknowledgements}

EC would like to thank the Lady Davis Fellowship Fund for financial support and
Yehuda Hoffman for the useful discussions.
He would also like to thank Roland Triay and Brent Tully for the invitation to the
Cosmic Flows 2016 at the $12^{th}$ \emph{Rencontres du Vietnam} conference where the idea behind this
work was first discussed. 
HAW is supported by the Beecroft Trust. 
The simulations used in this paper were performed on the NOTUR cluster \texttt{HEXAGON}, the computing facilities at the University of Bergen. 
DFM acknowledges support from the Research Council of Norway, and the NOTUR facilities.


\bibliographystyle{mn2e}
\bibliography{biblio}

\bsp

\appendix
\section{Resolution effects on the distributions of kinematic properties}
\label{appendix}

The distribution functions used throughout this \emph{paper} have been obtained using halo samples which included objects composed
by a number of particles as low as $\approx 40$.
In the present appendix we will show that the poor resolution of some haloes does not systematically bias our results, by 
comparing the changes induced to

\begin{itemize}
\item the number of objects as a function of the model
\item the distribution of \vtan
\item the distribution of \vrad
\item the distribution of total local group mass
\item the distribution of mass ratios
\end{itemize}

\noindent
by changing the particles' mass.
For this aim, we use a series of pure DM \LCDM\ simulations, within a 100\hMpc\ box and with Planck-I parameters, which 
were previously ran for testing purposes within the context of the Local Group Factory pipeline \citep{Carlesi:2016a}. 
We start with two $256^3$ particle simulations, SimuI and SimuII, with a mass resolution of $5.26\times10^{9}$\hMsun. For consistency reasons,
we introduce two different LG models (shown in \Tab{tab:lgmod}); in this way the smallest haloes in the sample will be
resolved with the same number of particles as the ones previously used.

\begin{table}
\begin{center}
\caption{Parameter intervals for the two Local Group models for the test simulations; \vrad\ and \vtan\ values are expressed
in \kms, $r$ in \hkpc\ and \mlg\ in $10^{12}$\hMsun\ units.}
\label{tab:lgmod}
\begin{tabular}{ccccc}
\hline
\quad & \vrad & \vtan & $r$ & \mlg\ \\
\hline
Model0 & [-125, -95] & [0, 500] & [0.44, 0.60] & [0.5, 2] \\
Model1 & [-500, -0] & [0, 500] & [0.25, 1.50] & [0.5, 7] \\
\hline
\end{tabular}
\end{center}
\end{table}

Then, for each simulation, we generate two sets of initial conditions with $512^3$ particles, 
for a total of four higher resolution realisations.
The first two of them share the same SimuI random phases on a $256^3$ grid, but use different 
phases for the generation of the white noise on the smaller scales. 
The second pair is generated in the same manner, on top of the SimuII white noise field.
This setting allows use to single out resolution effects from those induced by large-scale and small-scale cosmic variance.

\begin{figure*}
\begin{center}
\caption{Distribution functions for \vrad, \vtan\ and \mlg; each panel shows SimuI and SimuII (solid black lines) versus the
two higher-resolution re-simulations (a) (dotted red lines) and (b) (dashed blue lines). \label{app:img}}
$\begin{array}{cc}
\includegraphics[height=5.5cm]{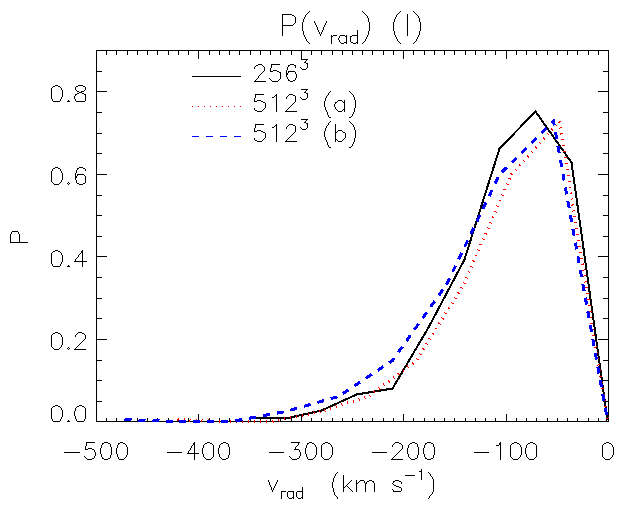} & 
\includegraphics[height=5.5cm]{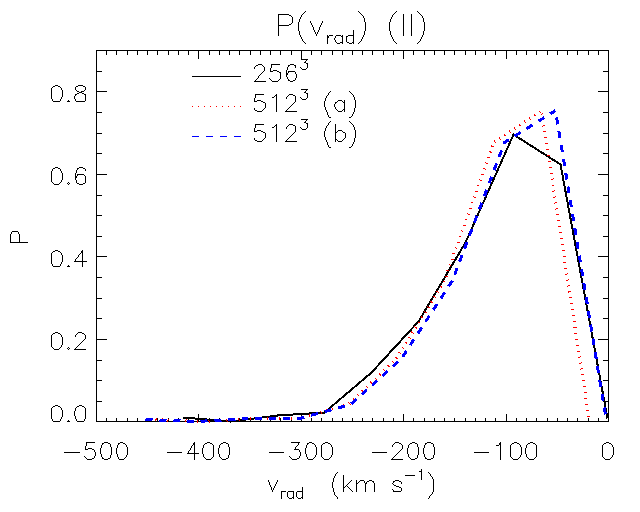} \\ 
\includegraphics[height=5.5cm]{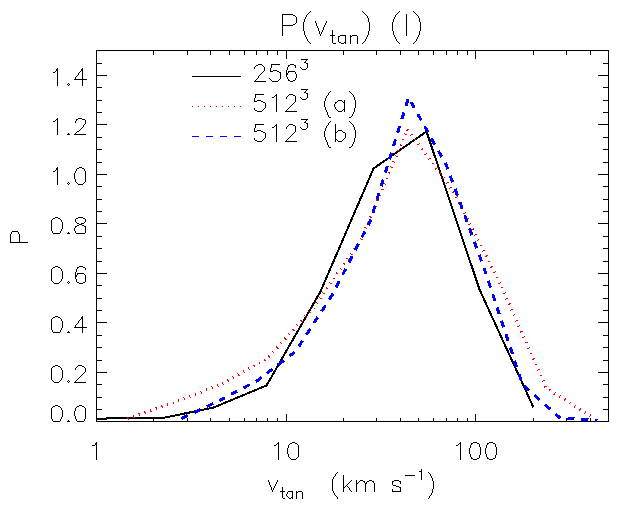} &
\includegraphics[height=5.5cm]{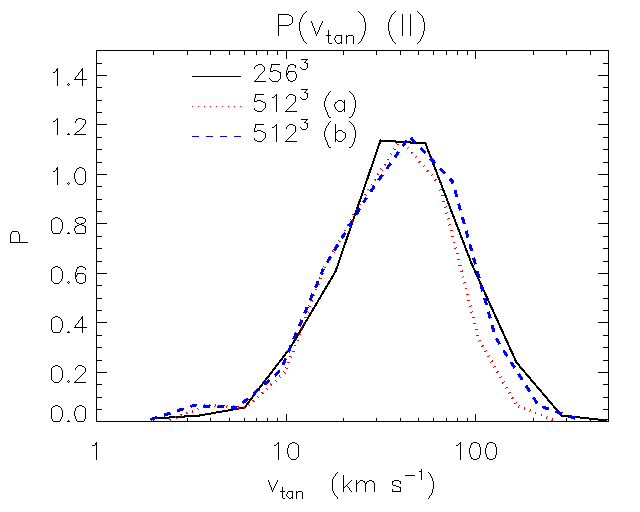} \\
\includegraphics[height=5.5cm]{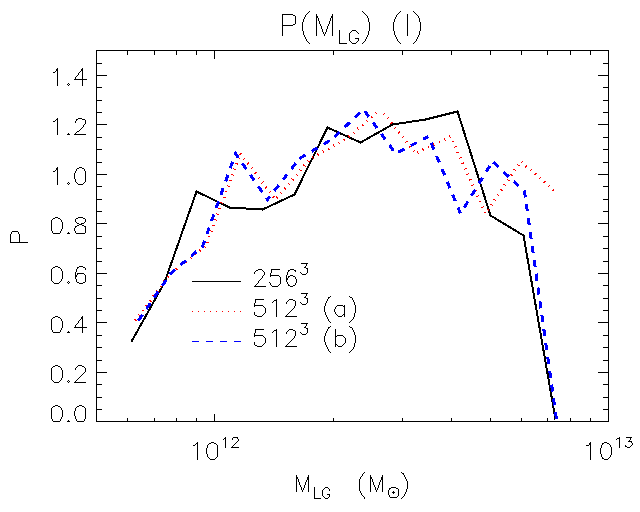} &
\includegraphics[height=5.5cm]{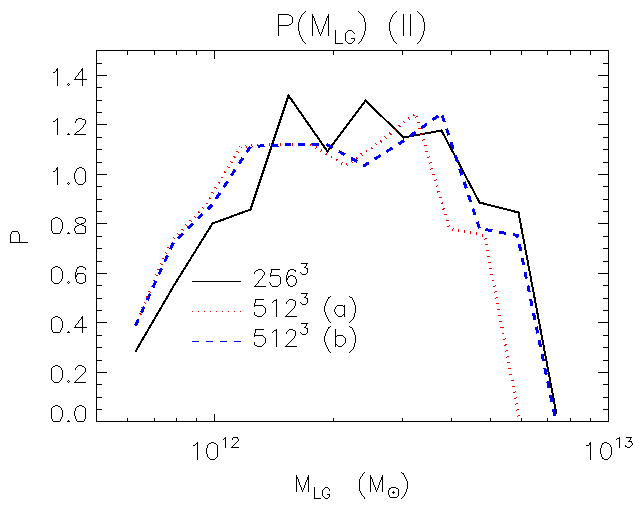} \\
\end{array}$
\end{center}
\end{figure*}

We use two Local Group Models, a restrictive one (Model0) and a second one with broader intervals (Model1), 
specified in \Tab{tab:mod}.
The first one will allow us to evaluate resolution effects on small halo samples, whereas 
with the second will provide us an estimate of their impact on the global distribution of LG variables.
We proceed computing distribution functions for masses and velocities and show the outcomes in 
\Tab{tab:res} and \Fig{app:img}.

In this work we study the dynamics of the Local Group (LG) within the context of cosmological models beyond General Relativity (GR).
Using observable kinematic quantities to identify candidate pairs we build up samples of simulated LG-like
  objects drawing from $f(R)$, symmetron, DGP and quintessence {\it N}-body simulations together with their \LCDM\ counterparts
featuring the same initial random phase realisations.
The variables and intervals used to define LG-like objects are referred to as Local Group model;
different models are used throughout this work and adapted to study their dynamical and kinematic properties.
The aim is to determine how well the observed LG-dynamics can be reproduced within cosmological theories beyond GR,
We compute kinematic properties of samples drawn from alternative theories and \LCDM\ and
compare them to actual observations of the LG mass, velocity and position.
As a consequence of the additional pull, pairwise tangential and radial velocities are enhanced in modified gravity and coupled dark energy with respect to \LCDM\,
inducing significant changes to the total angular momentum and energy of the LG.
For example, in models such as $f(R)$ and the symmetron
this increase can be as large as $60\%$, peaking well outside of the $95\%$ confidence region allowed by the data.
This shows how simple considerations about the LG dynamics can lead to clear small-scale observational signatures
for alternative scenarios, without the need of expensive high-resolution simulations.
\begin{table}
\begin{center}
\caption{Number of LG-like candidates as a function of model and realisation.}
\label{tab:mod}
\begin{tabular}{cccc}
\hline
$\quad$ & $256^3$ & $512^3 (a)$  & $512^3 (b)$ \\
\hline
N (SimuI, Model0) & 26 & 26 & 16 \\
N (SimuI, Model1) & 669 & 663 & 660 \\
N (SimuII, Model0)& 18 & 20 & 21 \\
N (SimuII, Model1) & 652 & 714 & 668 \\
\hline
\end{tabular}
\end{center}
\end{table}

\begin{table*}
\begin{center}
\caption{SimuI and SimuII results versus the two $512^3$ different short-wave
realisations of the same $256^3$ WN fields, using priors of Model1. \vrad\ and \vtan\ are
expressed in \kms.}
\begin{tabular}{cc}
\label{tab:res}
\begin{tabular}{cccc}
\hline
\multicolumn{4}{c}{SimuI}\\
\hline
$\quad$ & $256^3$ & $512^3 (a)$  & $512^3 (b)$ \\
\hline
\vrad\ & $ -57.85^{  26.32} _{ -29.74}$ & $ -52.60^{  24.36} _{ -39.27}$ & $ -49.96^{  23.01} _{ -36.93}$ \\
\vtan\ & $  45.39^{  26.35} _{ -17.31}$ & $  48.77^{  28.34} _{ -22.92}$ & $  42.04^{  28.12} _{ -16.59}$ \\
$\log_{10} M_{tot}$ & $  12.14^{   0.14} _{  -0.17}$ & $  12.13^{   0.14} _{  -0.15}$ & $  12.13^{   0.12} _{  -0.14}$\\
$M_{ratio}$ & $   1.58^{   0.82} _{  -0.38}$ & $   1.63^{   0.68} _{  -0.37}$ & $   1.56^{   0.74} _{  -0.35}$ \\
\hline
\end{tabular} & 

\begin{tabular}{ccc}
\hline
\multicolumn{3}{c}{SimuII}\\
\hline
$256^3$ & $512^3 (a)$  & $512^3 (b)$ \\
\hline
$ -55.62^{  22.85} _{ -31.65}$&$ -55.06^{  25.62} _{ -30.22}$&$ -51.36^{  22.69} _{ -32.56}$\\
$  44.73^{  32.14} _{ -16.67}$&$  45.08^{  28.88} _{ -21.98}$&$  42.78^{  30.54} _{ -19.46}$\\
$  12.15^{   0.13} _{  -0.16}$&$  12.13^{   0.13} _{  -0.15}$&$  12.12^{   0.14} _{  -0.14}$\\
$   1.70^{   0.61} _{  -0.45}$&$   1.57^{   0.68} _{  -0.31}$&$   1.62^{   0.67} _{  -0.38}$\\
\hline
\end{tabular}
\end{tabular}

\end{center}
\end{table*}

From these results we conclude the following:

\begin{itemize}
\item The shapes of all the distribution functions are largely unaffected by both cosmic variance and resolution.
\item The variance between different random realisations (both at the $256^3$ and at the $512^3$ level) is the largest source of differences
in the parameters distribution, as can be seen in the comparison between the two $256^3$ realisations (I) and (II) and also by looking 
at different $512^3$ simulations that share the same $256^3$ WN field. 
\item Increasing the resolution from $256^3$ to $512^3$ does not bias the distributions of masses, velocities and in 
the total number of objects-per-model found.
\end{itemize}

\noindent
These results show that the conclusions drawn in the present \emph{paper} are not affected by particle resolution, 
at least within the range of halo masses considered here.
Cosmic variance-related effects play a larger role, however, these are factored out from our results, since
the simulations have been compared on a same-seed basis.

\label{lastpage}

\end{document}